\documentclass[sigconf]{acmart}

\AtBeginDocument{%
  }

\setcopyright{acmlicensed}
\copyrightyear{2018}
\acmYear{2018}
\acmDOI{XXXXXXX.XXXXXXX}
\acmConference[Conference acronym 'XX]{Make sure to enter the correct
  conference title from your rights confirmation emai}{June 03--05,
  2018}{Woodstock, NY}
\acmISBN{978-1-4503-XXXX-X/18/06}

\definecolor{pblue}{rgb}{0.13,0.13,1}
\definecolor{pgreen}{rgb}{0,0.5,0}
\definecolor{pred}{rgb}{0.9,0,0}
\definecolor{pgrey}{rgb}{0.46,0.45,0.48}

\usepackage{listings}
\usepackage{tikz}
\usetikzlibrary{positioning,arrows.meta, decorations.pathreplacing, calc}

\begin{document}

\title{ICLF: An Immersive Code Learning Framework based on Git for Teaching and Evaluating Student Programming Projects}


\author{Pierre Schaus}
\email{pierre.schaus@uclouvain.be}
\orcid{0000-0002-3153-8941}
\affiliation{%
  \institution{UCLouvain}
  \city{Louvain-La-Neuve}
  \country{Belgium}
}

\author{Guillaume Derval}
\email{gderval@uliege.be}
\orcid{0000-0002-6700-3519}
\affiliation{%
  \institution{ULiège}
  \city{Liège}
  \country{Belgium}
}

\author{Augustin Delecluse}
\email{augustin.delecluse@uclouvain.be}
\orcid{0000-0001-6285-6515}
\affiliation{%
  \institution{UCLouvain}
  \city{Louvain-La-Neuve}
  \country{Belgium}
}

\renewcommand{\shortauthors}{Schaus et al.}

\begin{abstract}
Programming projects are essential in computer science education for bridging theory with practice and introducing students to tools like Git, IDEs, and debuggers. However, designing and evaluating these projects—especially in Massive Open Online Courses (MOOCs)—can be challenging. We propose
the Immersive Code Learning Framework (ICLF), a scalable Git-based organizational pipeline for managing and evaluating student programming project. Students begin with an existing code base, a practice that is crucial for mirroring real-world software development. Students then iteratively complete tasks that pass predefined tests. The instructor only manages a hidden parent repository containing solutions, which is used to generate an intermediate public repository with these solutions removed via a templating system. Students are invited collaborators on private forks of this intermediate repository, possibly updated throughout the semester whenever the teacher changes the parent repository. This approach reduces grading platform dependency, supports automated feedback, and allows the project to evolve without disrupting student work. Successfully tested over several years, including in an edX MOOC, this organizational pipeline provides transparent evaluation, plagiarism detection, and continuous progress tracking for each student.
\end{abstract}




\maketitle

\section{Introduction}

Programming projects are a fundamental component of students’ education to bridge theoretical knowledge with practical applications. 
Ideally during programming projects the students should also acquire a set important skills for a computer scientist like the ability to write clean code, test and debug it, integrate into an existing code bases, and collaborate with others using version controls systems. 
Unfortunately, many computer science programs still focus on theory or small projects starting from scratch, leaving graduates ill-equipped to work directly on more complex software projects.

Designing high-quality implementation projects and evaluating them effectively remains a significant challenge for educators, particularly in contexts such as Massive Open Online Courses (MOOCs), where a centralized manual correction is impractical.

Our motivation question is: \emph{How can we effectively teach students to work with large, real-world codebases while providing transparent, scalable, and automated evaluation in programming courses, particularly in contexts like MOOCs?}

To address the question we propose the \emph{Immersive Code Learning Framework} (ICLF) built around Git for teaching and evaluating student programming projects.
Instead of starting from scratch, students are placed in a scenario mirroring real-world software development.
Every student is invited as a collaborator on a git repository already containing code as a starting point.
This is the immersive aspect, where students engage with an existing code base rather than starting a project from scratch.
Students are then required to iteratively modify and extend this code base, completing tasks designed to meet predefined and graded unit tests.
The progress is thus continuously assessed, with students receiving feedback on their performance after each submission.

\begin{figure}[!ht]
    \centering
    \includegraphics[width=\linewidth]{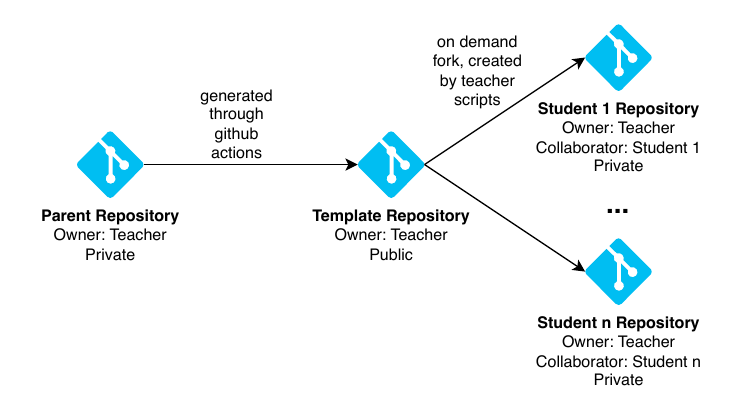}
    \caption{Set of repositories involved}
    \label{fig:repos}
\end{figure}

As illustrated on Figure~\ref{fig:repos}, the instructor owns the students’ repositories, which are private forks from another intermediate public repository that is itself derived from a hidden parent repository containing the original source code with solutions. 
The instructor only pushes changes to this parent repository. 
Whenever the instructor makes a push to the parent directly, the solutions are removed to generate the intermediate repository using a templating system embedded in the source code comments. 
Students are then invited to pull the updates from the intermediate repository to ensure their source code remains up-to-date.

ICLF minimizes the instructor’s dependency on a grading platform, ensuring that the code provided to students is well tested, functional, and the project remains feasible. 
Additionally, it allows instructors to dynamically evolve the project by adding tests or fixing bugs throughout the semester, even as students work on their initial tasks. 
The approach supports automated grading, plagiarism detection, and detailed tracking of individual progress without disrupting students’ workflow.

This methodology was successfully tested over several years, including in an edX MOOC.
Anonymous evaluations of our courses highlight that students appreciate this teaching process for its structured and transparent evaluation approach. They value the flexibility to run tests locally and, when confident in their code, to validate their progress within the secure grading environment.

ICLF is especially suitable for teaching related to systems development. 
This includes topics such as operating system development, compiler design, optimization, machine learning libraries, and more.

Our contribution is a detailed explanation of the framework. 
We also provide a small complete open-source instantiation of it for a java project so that anyone interested can reuse all the related scripts and source code.
We share our experience at using ICLF in two courses, including a MOOC.

Section \ref{sec:related_work}, gives the related work. 
Section \ref{sec:framework} gives a comprehensive explanation of the architecture, covering both functional and architectural aspects from the perspectives of students and teachers.  
Section \ref{sec:grading} then describes how the framework is used to grade exercises within a secured environment. 
Additionally, this section features a detailed "hello-world" open-source example of the framework using a Java library tailored for precise assessment of student code. 
In the final section \ref{sec:discussion}, we discuss our experiences using ICLF across various courses, including a MOOC on edX. This section also provides insights into how students engage with the framework.

\section{Related Work}
\label{sec:related_work}

Peer review in computer science education offers significant benefits for both students and educators~\cite{indriasari2020review}.
Regarding grading, students can be evaluated based on their peers’ assessments of their code. However, peer review alone has limitations, such as students struggling to determine whether their work meets expected goals and the delay in receiving feedback. These limitations are addressed by ICLF. In our view, peer review can serve as a complementary approach to ICLF.

In \emph{Inquiry-Based Learning} \cite{pedaste2015phases} learners actively engage through questioning, information gathering, and solution exploration. The goal is to encourage problem-solving skills. One such an approach is the so called \emph{creative problem solving} teaching method  \cite{van2014teaching} used for a MOOC on discrete optimization. 
In this approach, students are challenged to solve problems using any means necessary to achieve the desired outcome. Our approach differs somewhat, as students are tasked with integrating into an existing source code simulating the experience of joining a project in progress but, our approach is not mutually exclusive.

There are also similarities between our teaching framework and the one of \emph{Extreme Apprenticeship} (XA) \cite{vihavainen2012multi}. 
XA also allow students to run and execute tests locally as often as they wish. XA aims to guide students in using a working process similar to that of professional programmers. Our method mainly differs in that students interact with exercises through a git repository, rather than a plugin specifically developed in an IDE.
Also, XA does not automatically generate a template repository from the solution of the teacher and therefore need the so-called "Alpha-Beta-Open" release process to verify test thoroughness that our continuous deployment pipeline automates completely.

GitHub Classroom is a tool offered by GitHub sharing some similarity with our teaching framework and to some extent could be used to instantiate our framework. Students can join a classroom and work within a skeleton project. 
However, it does not allow them to collect automatically get the grades resulting from testing.
Additionally, there is no guarantee that the students have kept the provided unit tests intact, resulting in additional work from the teacher to validate the grades.

Web-CAT \cite{edwards2008web} is a Web tool for automated testing and grading of programming assignments. It is not Git-centric and requires students to upload their files manually or use a dedicated IDE plugin for submission. While Web-CAT is easier to use at the bachelor level, it is less aligned with the workflows of a professional development project compared to ICLF.

Autograder is a framework used to grade Java exercises \cite{helmick2007interface}. 
However, unlike JavaGrader, this tool does not rely on Junit5 \cite{junit5} standard testing library.
Other graders exist for Java using Junit \cite{kunchala2016java,gotel2007extending}, but these web-based tools are targeted for bachelors students.


\section{Description of the framework}
\label{sec:framework}

We explain in this section the teachers and students main interactions with the system. 
First, we discuss how students enroll in a course and begin working 
as a collaborator on a private fork of a intermediate template repository. 
Next, we explain how this public intermediate template repository is automatically generated from the private teacher parent repository using an operation called "stripping" removing and replacing parts of the source code.

We use GitHub as an example of a code hosting platform for the remainder of this paper;
other similar platforms can be used, such as GitLab or Bitbucket.
The only requirement for those platforms is to provide a rest API enabling users to create and interact with repositories \cite{github-rest-api} as well as a continuous delivery pipeline such as GitHub Actions to automate the execution of the scripts.

\subsection{Enrollment}
\label{sec:enrollment}

\begin{figure}[!ht]
    \centering
    	\begin{tikzpicture}[
		node distance=3cm and 1cm,
		font=\sffamily,
		lifeline/.style={draw, thick, minimum width=2.2cm, minimum height=0.5cm, align=center},
		dotted line/.style={densely dotted, thick, gray},
		msg/.style={-{Stealth}, thick},
		msgLabelUp/.style={above, anchor=south, align=center},
		msgLabelRight/.style={midway, right, anchor=west, align=left},
		msg dashed/.style={-{Stealth}, thick, dashed},
		fragment/.style={draw, thick, rounded corners},
		altlabel/.style={anchor=north west,
			font=\footnotesize\sffamily},
		guardlabel/.style={font=\footnotesize\itshape}
		]
		
		\node[lifeline] (student) {Student X};
		\draw[dotted line] (student.south) -- ++(0,-4.9);
		
		\node[lifeline, right of = student] (grading) {Grading System};
		\draw[dotted line] (grading.south) -- ++(0,-4.9);
		
		\node[lifeline, right of = grading] (template) {Template Repo};
		\draw[dotted line] (template.south) -- ++(0,-4.9);
		
		\draw[msg] ([yshift=-0.9cm]student.south) -- node[msgLabelUp]{Register for\\enrollment}
		([yshift=-0.9cm]grading.south);
		
		\draw[msg] ([yshift=-1.2cm]grading.south) -- ++ (0.5, 0) -- ++ (0, -0.5) node[msgLabelRight]{Create new\\repository for X} -- ++ (-0.5, 0);
		
		\draw[msg] ([yshift=-2.9cm]grading.south) -- node[msgLabelUp]{Retrieve template\\content}
		([yshift=-2.9cm]template.south);
		
		\draw[msg] ([yshift=-3.3cm]grading.south) -- ++ (0.5, 0) -- ++ (0, -0.5) node[msgLabelRight]{Initialize repo\\with template} -- ++ (-0.5, 0);
		
		\draw[msg] ([yshift=-4.8cm]grading.south) -- node[msgLabelUp]{Invite to collaborate\\ on new repo}
		([yshift=-4.8cm]student.south);
		
	\end{tikzpicture}
    \caption{Enrollment of a new student.}
    \label{fig:enrollment}
\end{figure}
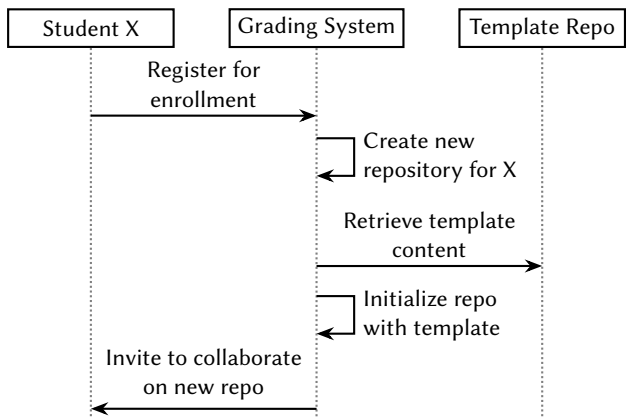

From the student's perspective\footnote{The description can easily be extended for groups of students working a same project and sharing the source code.}, enrolling in the course is simple - they only need to enter their GitHub username on a grading platform\footnote{(INGInious \cite{derval2015automatic} in our case).} and accept the invitation to collaborate on a private Git repository.
The git repository the student receives is a \textit{private} (i.e. not accessible to others, excepting the teaching staff, the owner of all the repositories) fork of template project, created by a GitHub bot account, that contains gaps representing missing implementations which they will need to complete over the course.
By preventing students from forking the template themselves and keeping the teacher as the owner, the teaching staff avoids solution sharing and ensures that students' repositories remain private.
The student is then ready to work and commit on its repository, filling the gaps and passing the graded unit tests.
Figure \ref{fig:enrollment} provides a sequence diagram of this process.

This methodology enables the teaching team to access all activities performed by the students on the repository. Since all repositories belong to the GitHub bot account owned by the teacher, it allows to retrieve all repositories and perform various tasks on them. This includes collecting analytics, check potential plagiarism issues as well as manually reviewing the code submitted by students.

\subsection{Student interactions on the repository}

The work performed by a student on its repository consists of filling the missing implementations, in an adequate order communicated by the teaching staff to the students.
Unit tests are provided for the missing implementations, allowing students to assess the correctness of their code locally.
Whenever an implementation has been successfully written by the student, the trainee can commit and push the changes on its repository. This is the only way for the student to test it fully, on the remote grading platform (we describe this process in section \ref{sec:grading}).
This has the benefit of forcing the student to interact with version control systems.

Moreover, the student must use git to interact with the template repository as well.
Indeed, the teacher can possibly update the template to fix a potential issue or to add another exercise to complete during the semester.
In such cases, the student must run git commands to integrate the changes from the template within its own project.
This results in merge operations, after which the student repository is up-to-date with respect to the template.

\subsection{Teacher update}

To generate the template repository, a teacher must have a project containing all solutions, with specific parts marked by delimiters to indicate which code needs to be stripped. The stripping operation involves removing the code written between these delimiters. An example of this operation is illustrated in Figure \ref{fig:stripping}. All the code between $\mathrm{BEGIN \ STRIP}$ and $\mathrm{END \ STRIP}$ comments disappears in the template. The code after $\mathrm{STUDENT}$ is injected to ensure that the student receives a code that compiles. The teachers can use any such delimiters in their project, which is minimally intrusive and ensures that they have a clear view of the level of difficulty of the exercises proposed to students.

\begin{figure}[!ht]
\begin{lstlisting}
public int increment(int x) {
    // STUDENT throw new RuntimeException("Not implemented");
    // BEGIN STRIP
    return x + 1;
    // END STRIP
}
\end{lstlisting}
\begin{lstlisting}
public int increment(int x) {
    throw new RuntimeException("Not implemented");
}
\end{lstlisting}
\caption{Example of a strip operation on a Java source file. 
The code on the top is the teacher (solution) implementation. 
The code below corresponds to the public template.
It is generated automatically by parsing the solution file and processing the strip tokens. We rely on a small utility tool for this:  \url{https://pypi.org/project/amanda/}}
\label{fig:stripping}
\end{figure}

The teacher can indicate which parts of the source code need to be stripped by adding tokens to the relevant files in the solution. This allows the teacher to have a clear overview of the gaps that need to be filled by the students. The combination of all the stripped files then creates the template repository, with gaps for the missing implementations that students must complete.

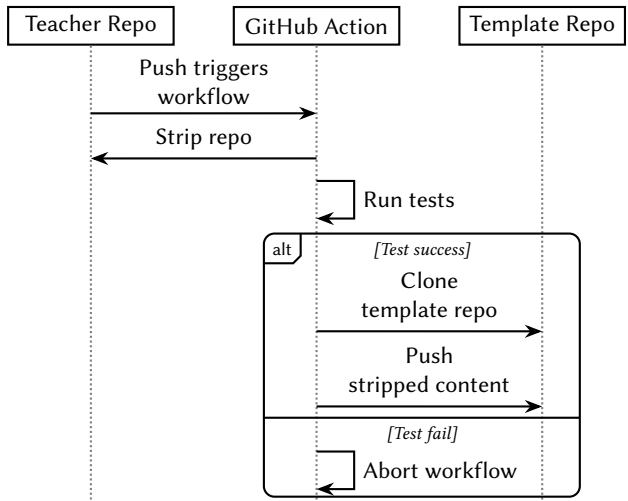
\begin{figure}[!ht]
    \centering
    	\begin{tikzpicture}[
	node distance=3cm and 1cm,
	font=\sffamily,
	lifeline/.style={draw, thick, minimum width=2.2cm, minimum height=0.5cm, align=center},
	dotted line/.style={densely dotted, thick, gray},
	msg/.style={-{Stealth}, thick},
	msgLabel/.style={above, anchor=south, align=center},
	msg dashed/.style={-{Stealth}, thick, dashed},
	fragment/.style={draw, thick, rounded corners},
	altlabel/.style={anchor=north west,
		font=\footnotesize\sffamily},
	guardlabel/.style={font=\footnotesize\itshape}
	]
	
	\node[lifeline] (teacher) {Teacher Repo};
	\draw[dotted line] (teacher.south) -- ++(0,-6.1);
	
	\node[lifeline, right of = teacher] (gha) {GitHub Action};
	\draw[dotted line] (gha.south) -- ++(0,-6.1);
	
	\node[lifeline, right of = gha] (template) {Template Repo};
	\draw[dotted line] (template.south) -- ++(0,-6.1);
	
	\draw[msg] ([yshift=-0.9cm]teacher.south) -- node[msgLabel]{Push triggers\\workflow}
	([yshift=-0.9cm]gha.south);
	
	\draw[msg] ([yshift=-1.5cm]gha.south) -- node[msgLabel]{Strip repo}
	([yshift=-1.5cm]teacher.south);
	
	\draw[msg] ([yshift=-1.8cm]gha.south) -- ++ (0.5, 0) -- ++ (0, -0.5) node[midway, right]{Run tests} -- ++ (-0.5, 0);
	
	\draw[msg] ([yshift=-3.8cm]gha.south) -- node[msgLabel]{Clone\\template repo}
	([yshift=-3.8cm]template.south);
	
	\draw[msg] ([yshift=-4.8cm]gha.south) -- node[msgLabel]{Push\\stripped content}
	([yshift=-4.8cm]template.south);
	
	\draw[msg] ([yshift=-5.4cm]gha.south) -- ++ (0.5, 0) -- ++ (0, -0.5) node[midway, right]{Abort workflow} -- ++ (-0.5, 0);
	
	\coordinate (altTopLeft) at ([yshift=-2.5cm, xshift=-0.7cm]gha.south);
	\coordinate (altBottomRight) at ([yshift=-6.0cm, xshift=0.5cm]template.south);
	
	\draw[fragment] (altTopLeft) rectangle (altBottomRight);
	
	\node[altlabel] (altClause) at (altTopLeft) {alt};
	\coordinate (almostSEfromSW) at ($(altClause.south east) + (-1mm,0)$);
	\coordinate (almostSEfromNE) at ($(altClause.south east) + (0,1mm)$);
	\draw[thick] (altClause.south west) -- (almostSEfromSW) -- (almostSEfromNE) -- (altClause.north east);

	\coordinate (splitLine) at ($(altTopLeft)!0.7!(altBottomRight)$); 
	\draw[thick] (altTopLeft|-splitLine) -- (altBottomRight|-splitLine);
	
	\coordinate (midPoint) at ($(altTopLeft)!0.5!(altBottomRight)$);
	\node[guardlabel, anchor=north] at (midPoint |- altTopLeft) {[Test success]};
	\node[guardlabel, anchor=north] at (midPoint |- splitLine) {[Test fail]};
	
\end{tikzpicture}
    \caption{The update of the template repository. 
    A GitHub action is triggered at each push on the solution repository, stripping the code and pushing it to the template project.}
    \label{fig:teacher-update}
\end{figure}

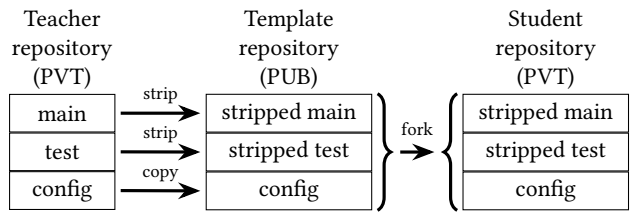
\begin{figure}[!ht]
    \centering
    	\begin{tikzpicture}[
    	node distance=1em and 2em,
    	>={Stealth[round]},
    	align=center,
    	myRectangle1/.style={
    		draw, 
    		rectangle, 
    		minimum width=1.4cm, 
    		minimum height=0.5cm
    	},
    	myRectangle2/.style={
    		draw, 
    		rectangle, 
    		minimum width=2.2cm, 
    		minimum height=0.5cm
    	},
    	description/.style={
    		align=center,
    	},
    	rightTo/.style={right=1.2cm of #1},
    	aboveTo/.style={above=-0.05cm of #1},
    	link/.style={
    		very thick,
    		-{Stealth[black]},
    		shorten >= 2pt, shorten <= 2pt
    	}
		]
		
		\node[myRectangle1] (mainT)   {main};
		\node[myRectangle1, below= 0cm of mainT] (testT)   {test};
		\node[myRectangle1, below= 0cm of testT] (configT) {config};
		
		\node[description, aboveTo=mainT] {Teacher\\repository\\(PVT)};
		
		\node[myRectangle2, rightTo=mainT] (mainTemp)   {stripped main};
		\node[myRectangle2, below= 0cm of mainTemp] (testTemp)   {stripped test};
		\node[myRectangle2, below= 0cm of testTemp] (configTemp) {config};
		
		\node[description, aboveTo=mainTemp] {Template\\repository\\(PUB)};
		
		\node[myRectangle2, rightTo=mainTemp] (mainS)   {stripped main};
		\node[myRectangle2, below= 0cm of mainS] (testS)   {stripped test};
		\node[myRectangle2, below= 0cm of testS] (configS) {config};
		
		\node[description, aboveTo=mainS] {Student\\repository\\(PVT)};
		
		\draw[link] (mainT.east)   -- node[above,font=\footnotesize] {strip} (mainTemp.west);
		\draw[link] (testT.east)   -- node[above,font=\footnotesize] {strip} (testTemp.west);
		\draw[link] (configT.east) -- node[above,font=\footnotesize] {copy}  (configTemp.west);
		
		
		\draw [decorate,decoration={brace,amplitude=5pt,raise=2pt}, very thick]
		(mainTemp.north east) -- (configTemp.south east) node[midway] (braceLeft) {};
		
		\draw [decorate,decoration={brace,amplitude=5pt,raise=2pt, mirror}, very thick]
		(mainS.north west) -- (configS.south west) node[midway] (braceRight) {};
		
		\draw[link, shorten >= 7pt, shorten <= 7pt] (braceLeft)   -- node[above=0.1cm,font=\footnotesize] {fork} (braceRight);
		
	\end{tikzpicture}
    \caption{Creation of the template repository through stripping and of the student repository through forking. Only the template repository is public, the teacher and student ones are private.}
    \label{fig:repo-creation}
\end{figure}

The continuous deployment capabilities of platforms like GitHub Actions or Gitlab CI/CD, enable
to automate entirely the generation and update of the template repository. 
Upon a push to the solution repository by the teacher, a script is executed to strip the solution and create a new version of the template repository. The updated version is then pushed to the template repository, making it instantly accessible to all users. Figure \ref{fig:teacher-update} illustrates this automated process. The relation between the teacher, the template and a student repository is represented on Figure \ref{fig:repo-creation}.

The teacher can also utilize the same approach to selectively hide certain unit tests from students. This feature enables a portion of the tests to be visible in the template, while keeping the rest hidden. 
The ability to hide tests can be useful in situations where a few simple tests are made available only for students to understand the expected behavior, while the remaining tests are hidden to prevent brute-forcing and encourage students to develop their own tests. 
One can also keep the tests consistent between the teacher and the template repository to ensure a fully transparent evaluation of student projects.

\section{Grading}
\label{sec:grading}



The unit tests in a project should provide a way to determine a student's grade when executed. However, certain modules in a project may be more crucial than others, and teachers need to be able to assign weights to relevant tests accordingly. To address this, we use JavaGrader, a Junit5 extension that we have developed that enables Java unit tests to output a graded report. Similar functionalities could be achieved with other programming languages such as Python or C++.
We also discuss how to set up an automated grading system that securely runs tests in a controlled environment—ensuring the original tests are execute, guaranteeing the validity of the final grade.

\subsection{Grading the tests}
\label{sec:java_grading}

To benefit from features introduced by Junit5 \cite{junit5} (such as parametric testing, parallel execution, tagging and filtering of the tests, etc.), we rely on an extension called JavaGrader. 
This extension adds additional functionalities to the tests in the form of annotations, similarly to python decorators. 
Here are the most important annotations provided by the extension:

\begin{itemize}
    \item \texttt{Grade} is the core annotation from JavaGrader. 
    It says that a unit test is graded and might contain optional parameters such as a maximum run time or a weight. 
    The maximum run time can be specified as either CPU timeout (how much time the current thread has been spent running the test) or as a wall-clock timeout (how much time has passed since the beginning of the test).
    \item \texttt{GradeFeedback} can be used to provide an additional message depending on the outcome of a unit test.
    \item \texttt{Forbid} prevents the use of a given library when executing the test.
    This is useful for exercises where a student is asked to write code without relying on a specific library or data structure.
    It is implemented by overriding the class loader from the thread running the test, ensuring that all classes loaded by the student are allowed.
\end{itemize}

An example usage can be found in Listing \ref{listing:javagrader}. 
More annotations are provided in the library. 
They cover some cases, such as conditional tests, that are only run if the previous was successful. 
This can be used for testing the time complexity of an algorithm, which should only be run if previous tests have first validated the correctness of the program.
JavaGrader can be used alongside other extensions when executing the test suite.

\begin{lstlisting}[caption={Transforming unit tests into graded tests by using JavaGrader. \texttt{mytest1} benefits from the forbidding of the \texttt{java.lang.Thread} library, and adds a cpu timeout on the test. Indeed, the cpu timeout would have been meaningless if the student run its code in a spawned thread, as it is not related to the initial thread running the test.},captionpos=b,label={listing:javagrader}]
$$@Grade
public class MyTests {
    $$@Test
    $$@Grade(value = 5, cpuTimeout = 1)
    $$@Forbid("java.lang.Thread")
    void mytest1() {
        //this works
        something();
    }
    $$@Test
    $$@Grade(value = 3)
    $$@GradeFeedback(message = "You forgot to consider this particular case [...]", on = FAIL)
    void mytest2() {
        //this doesn't
        somethingElse();
    }
}
\end{lstlisting}

\subsection{Student self-assessment}
\label{sec:running-test-locally}
As a fraction of tests from the teachers are provided in the template (and thus in the student) repository, the trainees can use them to their advantage.
They can locally run (a part of) the graded tests, and obtain an immediate feedback, along with debugging possibilities, directly on their computer, even when offline. 
The students are then able to iterate more efficiently on their code; this reduces both the overall effort of the code-evaluate-fix loop and the computation load on the online grading system. 
More advanced tests, requiring the usage of a secret or specific computing capabilities not present on trainees' computers, can be stripped and made available on the online grading system only (as detailed in the next section).

\subsection{Running the tests in a secure environment}
\label{sec:running-test-on-INGInious}

Truly assessing the grades from the students requires particular caution. 
Simply retrieving a student repository and running its tests does not guarantee that their output is valid: nothing prevents a student from pushing a modified version of the tests. 
Furthermore, it might be a good practice to let the students modify the tests: they should be free of adding more tests cases if they wish, to convince themselves of the correctness of their code.

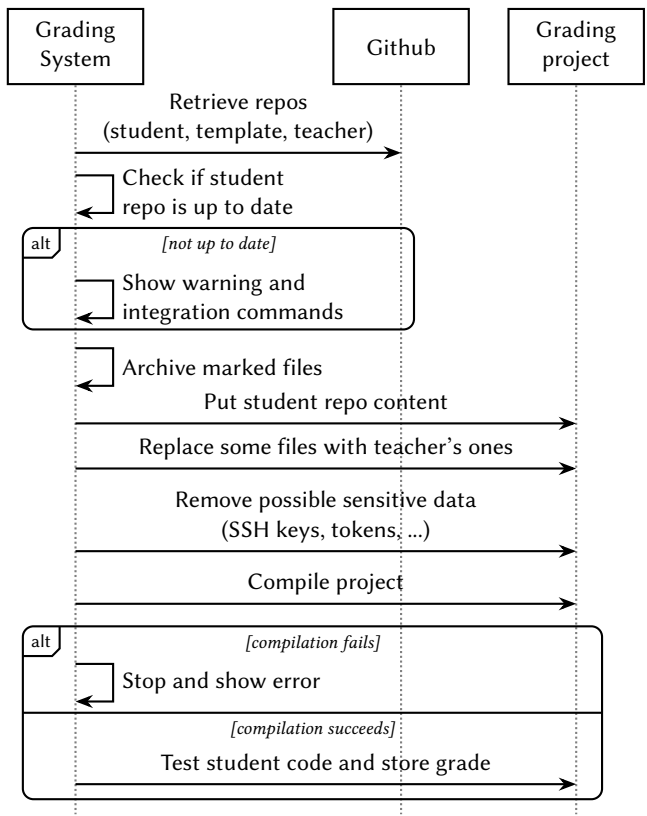
\begin{figure}
    \centering
    	\begin{tikzpicture}[
	font=\sffamily,
	lifeline/.style={draw, thick, minimum width=1.8cm, minimum height=1cm, align=center},
	dotted line/.style={densely dotted, thick, gray},
	msg/.style={-{Stealth}, thick},
	msgLabelUp/.style={above, anchor=south, align=center},
	msgLabelRight/.style={midway, right, anchor=west, align=left},
	msg dashed/.style={-{Stealth}, thick, dashed},
	fragment/.style={draw, thick, rounded corners},
	altlabel/.style={anchor=north west,
		font=\footnotesize\sffamily},
	guardlabel/.style={font=\footnotesize\itshape}
	]
	
	\node[lifeline] (system) {Grading\\System};
	\node[lifeline, right = 2.5cm of system] (github) {Github};
	\node[lifeline, right = 0.5cm of github] (project) {Grading\\project};
	\draw[dotted line] (system.south) -- ++(0,-9.7);
	\draw[dotted line] (github.south) -- ++(0,-9.7);
	\draw[dotted line] (project.south) -- ++(0,-9.7);
	
	\draw[msg] ([yshift=-0.9cm]system.south) -- node[msgLabelUp]{Retrieve repos\\(student, template, teacher)}
	([yshift=-0.9cm]github.south);
	
	\draw[msg] ([yshift=-1.2cm]system.south) -- ++ (0.5, 0) -- ++ (0, -0.5) node[msgLabelRight]{Check if student\\repo is up to date} -- ++ (-0.5, 0);
	
	
	\coordinate (altUpToDateTopLeft) at ([yshift=-1.9cm, xshift=-0.7cm]system.south);
	\coordinate (altUpToDateBottomRight) at ([yshift=-3.25cm, xshift=4.5cm]system.south);
	
	\draw[fragment] (altUpToDateTopLeft) rectangle (altUpToDateBottomRight);
	
	\node[altlabel] (altUpToDateClause) at (altUpToDateTopLeft) {alt};
	\coordinate (almostSEfromSWUpToDate) at ($(altUpToDateClause.south east) + (-1mm,0)$);
	\coordinate (almostSEfromNEUpToDate) at ($(altUpToDateClause.south east) + (0,1mm)$);
	\draw[thick] (altUpToDateClause.south west) -- (almostSEfromSWUpToDate) -- (almostSEfromNEUpToDate) -- (altUpToDateClause.north east);
	
	\coordinate (midPointaltUpToDate) at ($(altUpToDateTopLeft)!0.5!(altUpToDateBottomRight)$);
	\node[guardlabel, anchor=north] at (midPointaltUpToDate |- altUpToDateTopLeft) {[not up to date]};
	
	\draw[msg] ([yshift=-2.6cm]system.south) -- ++ (0.5, 0) -- ++ (0, -0.5) node[msgLabelRight]{Show warning and\\integration commands} -- ++ (-0.5, 0);
	
	
	\draw[msg] ([yshift=-3.5cm]system.south) -- ++ (0.5, 0) -- ++ (0, -0.5) node[msgLabelRight]{Archive marked files} -- ++ (-0.5, 0);
	
	\draw[msg] ([yshift=-4.5cm]system.south) -- node[msgLabelUp]{Put student repo content}
	([yshift=-4.5cm]project.south);
	
	\draw[msg] ([yshift=-5.1cm]system.south) -- node[msgLabelUp]{Replace some files with teacher's ones}
	([yshift=-5.1cm]project.south);
	
	\draw[msg] ([yshift=-6.2cm]system.south) -- node[msgLabelUp]{Remove possible sensitive data\\(SSH keys, tokens, ...)}
	([yshift=-6.2cm]project.south);
	
	\draw[msg] ([yshift=-6.9cm]system.south) -- node[msgLabelUp]{Compile project}
	([yshift=-6.9cm]project.south);
	
	
	\coordinate (altGradingTopLeft) at ([yshift=-7.2cm, xshift=-0.7cm]system.south);
	\coordinate (altGradingBottomRight) at ([yshift=-9.5cm, xshift=7cm]system.south);
	
	\draw[fragment] (altGradingTopLeft) rectangle (altGradingBottomRight);
	
	\node[altlabel] (altGradingClause) at (altGradingTopLeft) {alt};
	\coordinate (almostSEfromSWGrading) at ($(altGradingClause.south east) + (-1mm,0)$);
	\coordinate (almostSEfromNEGrading) at ($(altGradingClause.south east) + (0,1mm)$);
	\draw[thick] (altGradingClause.south west) -- (almostSEfromSWGrading) -- (almostSEfromNEGrading) -- (altGradingClause.north east);
	
	\coordinate (splitLineGrading) at ($(altGradingTopLeft)!0.5!(altGradingBottomRight)$); 
	\draw[thick] (altGradingTopLeft|-splitLineGrading) -- (altGradingBottomRight|-splitLineGrading);
	
	\coordinate (midPointGrading) at ($(altGradingTopLeft)!0.5!(altGradingBottomRight)$);
	\node[guardlabel, anchor=north] at (midPointGrading |- altGradingTopLeft) {[compilation fails]};
	
	\draw[msg] ([yshift=-7.7cm]system.south) -- ++ (0.5, 0) -- ++ (0, -0.5) node[msgLabelRight]{Stop and show error} -- ++ (-0.5, 0);
	
	\node[guardlabel, anchor=north] at (midPointGrading |- splitLineGrading) {[compilation succeeds]};
	
	\draw[msg] ([yshift=-9.3cm]system.south) -- node[msgLabelUp]{Test student code and store grade}
	([yshift=-9.3cm]project.south);
	
	
\end{tikzpicture}
    \caption{Grading a task within a grading system. The feedback is constructed by the grading system and contains the information related to the grading and notes for the students.}
    \label{fig:run-grading}
\end{figure}
Figure \ref{fig:run-grading} depicts the sequence diagram for grading a task, with the key steps outlined next.
\begin{enumerate}
    \item \label{grading:retrieve} The student, template and teacher repositories are retrieved.
    \item \label{grading:check-commit} It is ensured that the last commit from the template appears within the student repository. If not, a warning will be given as feedback letting the student know that the repository need to be up-to date to be graded. Some commands are prompted to help the student integrate the changes from the template.
    \item \label{grading:archive} Some marked files and directories from the student repository are archived. The archived files can be used for further analysis by the teaching team such as plagiarism detection between students with a tool like JPlag \cite{prechelt2000jplag}.
    \item \label{grading:replace} The grading project is prepared by replacing files or directories from the student to put the ones from the teacher instead (typically the tests, including hidden ones, and configuration but possibly additional files such as a sensitive dataset that must be hidden to students).
    \item \label{grading:remove-sensible} The sensible files that are mandatory for cloning the repositories are removed (ssh keys, token files, etc.) 
    \item \label{grading:compile} The grading project is compiled. Failure to compile them stops the evaluation with an error message corresponding to the error printed by the compiler.
    \item \label{grading:run-test} Test the student code and give feedback to the student. The grades are stored.
\end{enumerate}
To prevent any memory corruption and fully automate the grading, we use a grading platform such as INGInious \cite{derval2015automatic}\footnote{Other plateforms such as GradeScope \citep{singh2017gradescope} could also be used.}. 
INGInious essentially run these steps in a \textit{jailed} environment ensuring, among other things, that the code cannot interact with anything other than the grader, cannot access the internet, and cannot read or write files on the file system except where they are specifically allowed to.
This platform is also used with a special task for creating the repository, using the methodology presented in section \ref{sec:enrollment}.



\begin{figure}[!ht]
    \centering
    \includegraphics[width=\linewidth]{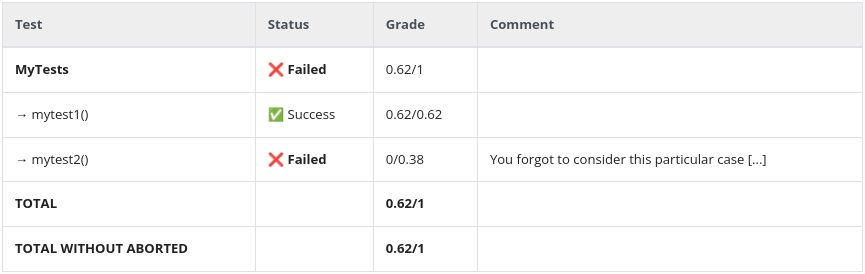}
    \caption{Output of the tests from Listing \ref{listing:javagrader}. As the class \texttt{MyTests} is also annotated with \texttt{@Grade} without specifying optional parameters, the maximum grade is 1. This value can be overridden by changing the class annotation.}
    \label{fig:rst-screenshot}
\end{figure}

The feedback given back to the student in step \ref{grading:run-test} consists of the grade for the task, 
and any possible information proving useful feedback to the student.
In the case of JavaGrader that we use, an example output related to the tests from Listing \ref{listing:javagrader}, is presented on Figure \ref{fig:rst-screenshot}.

Setting up a grading task can be done in a matter of minutes.
Except for the homework statement, only a few parts need to change between two grading tasks. 
The procedure remains the same across the tasks for one course, with the exception of steps \ref{grading:archive} and \ref{grading:replace}, where different files or directory can be specified, and for the tests that needs to be run.

As an example, we provide a detailed "hello-world" open-source implementation of the whole pipeline.
The example is accessible through \url{https://inginious.org/course/iclf-example}. 
It uses INGInious as the grading system. The related repositories can be found at \url{https://github.com/OneAnonymizedUser/teacher-repository} for the teacher repository and at \url{https://github.com/OneAnonymizedUser/template-repository} for the template. For demonstrating purposes, the teacher repository is public, but it should remain private in a real scenario. 

\section{Use cases}
\label{sec:discussion}

The teaching framework presented in this paper was successfully used in two courses for more that three years: A Discrete Optimization Course and a Constraint Programming Course on edX.

\subsection{A Discrete Optimization Course}

This course focuses on advanced algorithms for discrete optimization, with theoretical content similar to that of \cite{van2014teaching}.
Each task is related to a different optimization technique, one of these is the well-known branch-and-bound method.
Students are provided with code for cumbersome tasks such as instance parsing, along with an incomplete skeleton of a generic algorithm (e.g., branch-and-bound) that they need to complete and then instantiate on a problem (e.g., the traveling salesman problem).
Some of the graded tests are hidden from the students, primarily the most challenging instances, making it difficult to achieve the maximum grade, in line with the method introduced in \cite{van2014teaching}.
The complete teaching framework, as described in this paper, was utilized over three years, with about 100 students each year.

Each year, our teaching team prepared the projects just in time. 
As a result, students had to pull the next project, which appeared as a new package in the template source code, every two weeks.
Anonymous feedback from students demonstrated their appreciation for the teaching framework and course format.

\begin{figure}[!ht]
    \centering
    \includegraphics[width=\linewidth]{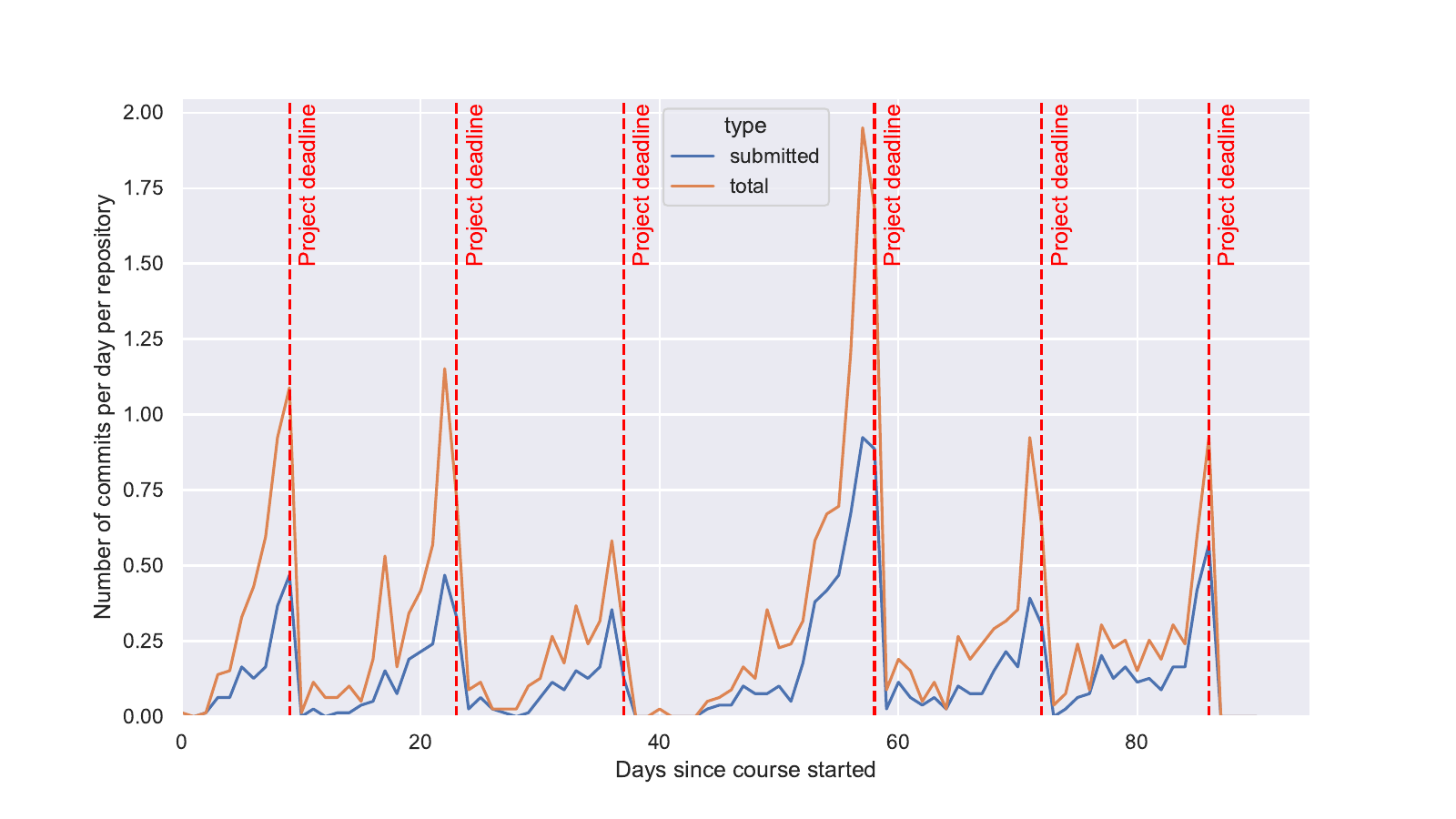}
    \caption{Average number of commits (per student-repository) over time for the Discrete Optimization course. We report the total number of commits and the number of commits that were submitted to the grading system. The commits related to merges from the template, fetching the new projects instructions, are omitted.
    }
    \label{fig:linfo2266-commits-over-time}
\end{figure}

As the framework enables the collection of global statistics, we computed the average number of commits over time by the students in a typical year, as shown in Figure \ref{fig:linfo2266-commits-over-time} to get insights on how and when students interact with the framework.
We observe that, as expected, students tend to work closer to deadlines, with a peak in the number of commits before each due date.
Interestingly we observed that not all commits correspond to a grading request, indicating that students prefer to assess their code locally before submitting their work.

\subsection{MOOC on Constraint Programming on edX}
\label{sec:discussion-cp}

We have taught a Constraint Programming (CP) course on edX for three years to an audience of approximately 300 students per year (100 students from our university, and 200 online external students) using the proposed methodology\footnote{Inginious enables the Learning Tool Interoperability (LTI) connection with edX}.
Most university courses on constraint programming teach it at the user level of a given library.
In contrast, ICLF enabled us to have students take charge of the development of an entire constraint programming solver library.
For students, a basic constraint programming solver is a fairly large piece of software, consisting of around 10,000 lines of code, excluding tests.
The sequence of proposed tasks starts at the lower levels of the solver and gradually moves toward applications, ensuring there is no “dark magic” for students from beginning to end.
There are no fixed deadlines for the projects, but students are strongly encouraged to stay up-to-date with the exercises, as some require code and algorithms from earlier projects to complete.
Although not included here due to space constraints, we collected data on the commits made by students each day during the semester.
We observed a higher number of commits on Fridays, corresponding to on-site practice sessions organized at our university.
Additionally, we noted that the number of commits gradually increased as the semester progressed, peaking as the final deadline approached.

\section{Conclusion}
\label{sec:conclusion}

We have presented the Immersive Code Learning Framework (ICLF), a scalable, and effective approach
based on git to teaching and grading programming projects.
By placing students in an environment with pre-existing codebases, the framework simulated real-world software development scenarios, fostering skills such as debugging, version control, and iterative development.

The framework’s ability to provide instant feedback, automate grading, and dynamically evolve projects was successfully demonstrated in diverse educational settings, including a Discrete Optimization course and a Constraint Programming MOOC. 
Anonymous evaluations from students underscored the value of ICLF in offering a structured, transparent, and rewarding learning process.

\begin{acks}
Augustin Delecluse is supported by Service Public de Wallonie Recherche under grant n°2010235 – ARIAC by DIGITALWALLONIA4.A
\end{acks}

\bibliographystyle{ACM-Reference-Format}
\bibliography{biblio}

@article{derval2015automatic,
  title={Automatic grading of programming exercises in a MOOC using the INGInious platform},
  author={Derval, Guillaume and Gego, Anthony and Reinbold, Pierre and Frantzen, Benjamin and Van Roy, Peter},
  journal={European Stakeholder Summit on experiences and best practices in and around MOOCs (EMOOCS’15)},
  pages={86--91},
  year={2015}
}

@inproceedings{edwards2008web,
  title={Web-CAT: automatically grading programming assignments},
  author={Edwards, Stephen H and Perez-Quinones, Manuel A},
  booktitle={Proceedings of the 13th annual conference on Innovation and technology in computer science education},
  pages={328--328},
  year={2008}
}

@article{indriasari2020review,
  title={A review of peer code review in higher education},
  author={Indriasari, Theresia Devi and Luxton-Reilly, Andrew and Denny, Paul},
  journal={ACM Transactions on Computing Education (TOCE)},
  volume={20},
  number={3},
  pages={1--25},
  year={2020},
  publisher={ACM New York, NY, USA}
}

@book{prechelt2000jplag,
  title={JPlag: Finding plagiarisms among a set of programs},
  author={Prechelt, Lutz and Malpohl, Guido and Philippsen, Michael},
  year={2000},
  publisher={Univ., Fak. f{\"u}r Informatik}
}

@inproceedings{van2014teaching,
  title={Teaching creative problem solving in a MOOC},
  author={Van Hentenryck, Pascal and Coffrin, Carleton},
  booktitle={Proceedings of the 45th ACM technical symposium on Computer science education},
  pages={677--682},
  year={2014}
}

@inproceedings{vihavainen2012multi,
  title={Multi-faceted support for MOOC in programming},
  author={Vihavainen, Arto and Luukkainen, Matti and Kurhila, Jaakko},
  booktitle={Proceedings of the 13th annual conference on Information technology education},
  pages={171--176},
  year={2012}
}

@article{pedaste2015phases,
  title={Phases of inquiry-based learning: Definitions and the inquiry cycle},
  author={Pedaste, Margus and M{\"a}eots, Mario and Siiman, Leo A and De Jong, Ton and Van Riesen, Siswa AN and Kamp, Ellen T and Manoli, Constantinos C and Zacharia, Zacharias C and Tsourlidaki, Eleftheria},
  journal={Educational research review},
  volume={14},
  pages={47--61},
  year={2015},
  publisher={Elsevier}
}

@misc{junit5,
  title = {{JUnit 5}},
  year = {2023},
  month = mar,
  note = {[Online; accessed 21. Mar. 2023]},
  url = {https://junit.org/junit5}
}

@inproceedings{singh2017gradescope,
  title={Gradescope: a fast, flexible, and fair system for scalable assessment of handwritten work},
  author={Singh, Arjun and Karayev, Sergey and Gutowski, Kevin and Abbeel, Pieter},
  booktitle={Proceedings of the fourth (2017) acm conference on learning@ scale},
  pages={81--88},
  year={2017}
}

@inproceedings{helmick2007interface,
  title={Interface-based programming assignments and automatic grading of java programs},
  author={Helmick, Michael T},
  booktitle={Proceedings of the 12th annual SIGCSE conference on Innovation and technology in computer science education},
  pages={63--67},
  year={2007}
}

@article{kunchala2016java,
  title={Java auto grader},
  author={Kunchala, Ashrita and Rao, Maneesh Gunnala Ranga},
  year={2016}
}

@inproceedings{gotel2007extending,
  title={Extending and contributing to an open source web-based system for the assessment of programming problems},
  author={Gotel, Olly and Scharff, Christelle and Wildenberg, Andy},
  booktitle={Proceedings of the 5th international symposium on Principles and practice of programming in Java},
  pages={3--12},
  year={2007}
}

@misc{github-rest-api,
  title = {{GitHub REST API documentation - GitHub Docs}},
  journal = {GitHub Docs},
  year = {2023},
  month = mar,
  note = {[Online; accessed 22. Mar. 2023]},
  url = {https://docs.github.com/en/rest}
}


\end{document}